\documentclass[journal=jacsat,manuscript=article]{achemso}

\usepackage{chemformula} 
\usepackage[T1]{fontenc} 
\usepackage{physics}
\usepackage{braket}
\newcommand*{\citen}{}
\DeclareRobustCommand*{\citen}[1]{%
  \begingroup
    \romannumeral-`\x 
    \setcitestyle{numbers}%
    \cite{#1}%
  \endgroup
}

\author{Huiyao Chen}
\affiliation[Cornell]
{Cornell University, Ithaca, New York 14853, United States}
\author{Noah F. Opondo}
\affiliation[Purdue]
{Purdue University, West Lafayette, Indiana 47907, United States}
\author{Boyang Jiang}
\affiliation[Purdue]
{Purdue University, West Lafayette, Indiana 47907, United States}
\author{Evan R. MacQuarrie}
\affiliation[Cornell]
{Cornell University, Ithaca, New York 14853, United States}
\author{Rapha$\ddot{\text{e}}$l S. Daveau}
\affiliation[Cornell]
{Cornell University, Ithaca, New York 14853, United States}
\author{Sunil A. Bhave}
\affiliation[Purdue]
{Purdue University, West Lafayette, Indiana 47907, United States}
\author{Gregory D. Fuchs}
\email{gdf9@cornell.edu}
\affiliation[Cornell]
{Cornell University, Ithaca, New York 14853, United States}
\alsoaffiliation[Kavli]
{Kavli Institute at Cornell for Nanoscale Science, Cornell University, Ithaca, New York 14853, United States}

\title[SILpaper]
  {Engineering electron-phonon coupling of quantum defects to a semi-confocal acoustic resonator}

\abbreviations{NV, MEMS, SiC}

\begin{document}


\begin{abstract}
 Diamond-based microelectromechanical systems (MEMS) enable direct coupling between the quantum states of nitrogen-vacancy (NV) centers and the phonon modes of a mechanical resonator. One example, diamond high-overtone bulk acoustic resonators (HBARs), feature an integrated piezoelectric transducer and support high-quality factor resonance modes into the GHz frequency range. The acoustic modes allow mechanical manipulation of deeply embedded NV centers with long spin and orbital coherence times. Unfortunately, the spin-phonon coupling rate is limited by the large resonator size, $>100~\mu$m, and thus strongly-coupled NV electron-phonon interactions remain out of reach in current diamond BAR devices. Here, we report the design and fabrication of a semi-confocal HBAR (SCHBAR) device on diamond (silicon carbide) with $f\cdot Q>10^{12}$($>10^{13}$). The semi-confocal geometry confines the phonon mode laterally below 10~$\mu$m. This drastic reduction in modal volume enhances defect center electron-phonon coupling. For the native NV centers inside the diamond device, we demonstrate mechanically driven spin transitions and show a high strain-driving efficiency with a Rabi frequency of $(2\pi)2.19(14)$~MHz/V$_{p}$, which is comparable to a typical microwave antenna at the same microwave power.  
\end{abstract}

Defect-based qubits are attractive platforms for solid state quantum technologies\cite{awschalom2018quantum}. The leading examples are the nitrogen-vacancy (NV)\cite{doherty2013nitrogen} center and the silicon-vacancy (SiV)\cite{sukachev2017silicon} center in diamond, and the divacancy center\cite{christle2015isolated} and the silicon vacancy center (V$_{\text{Si}}$)\cite{widmann2015coherent} in silicon carbide (SiC). Hybrid quantum systems based on these defect qubits are particularly interesting because they interface the qubit spin to photons or phonons and
thus potentially enable the transport of quantum information. For sensing applications, they offer unconventional modalities of quantum control which is a resource for extending the coherence time and thus sensitivity.  Coupling spins to mechanical motion could also enable new quantum-enhanced sensors of motion, such as inertial sensing\cite{ajoy2012stable,ledbetter2012gyroscopes}.

Although solid state spin-photon entanglement has been demonstrated in recent years\cite{bernien2013heralded} and has been used to build quantum networks\cite{kalb2017entanglement}, defect-based spin-mechanical systems have yet to operate at the single phonon quantum level because they are limited by weak electron-phonon coupling, $g$, in existing devices. Considering $g\propto\sqrt{1/V}$, where $V$ is the modal volume, one approach to strengthening the coupling is to engineer small mode volume mechanical resonators with high quality factors. Ultimately, defect-based spin-mechanical systems may enable new sensing applications and control of phonon states at the quantum level\cite{o2010quantum}.


Defect-based spin-mechanical systems can be classified into two categories: 1) micro-beam resonator systems\cite{ovartchaiyapong2014dynamic,burek2016diamond,cady2019diamond} and 2) micro-electromechanical systems (MEMS)\cite{macquarrie2013mechanical,golter2016optomechanical,chen2018orbital,whiteley2019spin} with integrated thin-film piezoelectric transducers. While the first category minimizes the resonator fabrication to a single material, i.e., diamond, SiC, etc., high-frequency micro-beam resonators ($>$1~GHz) are difficult to efficiently excite and detect. The latter category includes surface acoustic wave (SAW) and bulk acoustic wave (BAW) devices. These resonator types can be excited and characterized electrically, and they enable both quantum circuit integration \cite{o2010quantum, chu2017quantum} and direct quantum control of embedded defect qubits. At higher frequency ($>$2~GHz), while SAW devices start to show progressively higher losses, BAW resonators maintain a high quality factor\cite{aigner2008saw}. Additionally, BAW resonator can couple to defect centers that lie deep inside the device, making them less susceptible to deleterious surface effects and thus they can possess long spin and orbital coherence times.

Here we report the design, fabrication and performance of a new type of diamond (and SiC) BAW device, the semi-confocal high-overtone bulk acoustic resonator (SCHBAR). The device features a micro-scale phonon mode volume and integrated atomic-scale quantum defects. Electrical measurements show that the frequency-quality factor product is $f\cdot Q>10^{12}$ ($>10^{13}$) for diamond (SiC) SCHBAR devices operating with a $\sim$3~GHz phonon mode. In a diamond SCHBAR device, we use the native NV centers to characterize the a.c. strain performance of the acoustic resonator. We mechanically drive NV center ground-state spin Rabi oscillations at a rate $(2\pi)2.19(14)$~MHz/V$_{p}$, corresponding to a mechanical strain $1.59(14)\times10^{-4}~\text{V}_{p}^{-1}$, indicating high power-to-strain conversion efficiency. Applications of this device include the fast quantum control of defect spins in the double-quantum basis, study of spin dynamics in the strong-mechanical-driving regime, and quantum control of resonator states using dense defect spin ensembles.


BAW devices have been widely used in the microelectronics industry for RF filtering and other wireless applications\cite{bhugra2017piezoelectric}. The transducer of most BAW devices consists of a piezoelectric thin film sandwiched between two electrodes in a released structure or solidly mounted on a planar substrate. The two planar boundaries of the device form a cavity for the bulk acoustic wave. To achieve a high quality factor, a high degree of parallelism for the two planar boundaries is stringently required to suppresses the acoustic diffraction loss\cite{baron2013high}. Even with perfectly parallel boundaries, a planar cavity lacks lateral confinement\cite{siegman1986lasers}. A partial solution is to make large area transducer to spatially limit the diffraction loss, however, this also constrains the size scale of a standard BAW device to be larger than $100~\mu$m. 

In analogy to a confocal or semi-confocal optical cavity, an acoustic cavity with curved boundaries that match the wavefront of the acoustic wave can overcome diffraction loss and provide stable mode confinement. The cavity stability criterion suggests that the best confinement is obtained in a confocal geometry or a semi-confocal geometry for a planar convex boundary condition. The idea has been demonstrated recently in both SAW\cite{fang1989saw, shilton2008particle, whiteley2019spin} and BAW devices\cite{galliou2013extremely,kharel2018ultra} at millimeter scale. Here we apply this concept to a micro-scale BAW resonator. We choose to work with diamond and SiC substrates because they possess excellent mechanical properties and their crystal lattices host quantum defects. We design the micro resonator for a 10- and a 20-$\mu$m-thick device.

\begin{figure}
\includegraphics{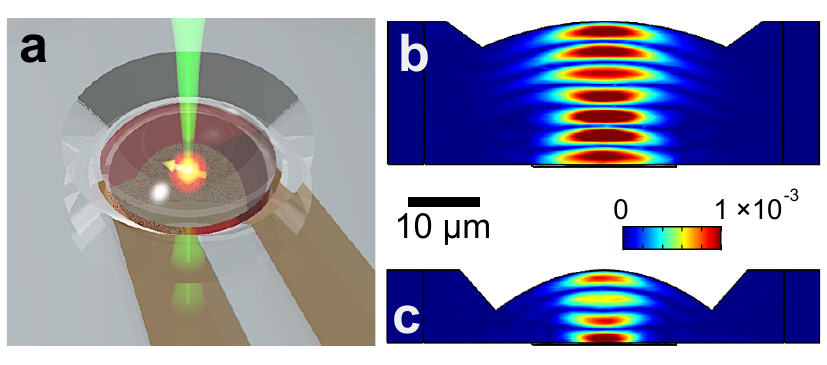}
\caption{(a) Concept image of the semi-confocal HBAR device design. Solid state defect spins (orange arrow) such as from diamond NV centers are accommodated in the depth of the substrate. They can be addressed optically (laser), magnetically (microwave) and mechanically (acoustics). Brown-colored leads represent the transducer electrodes. The ruby-colored layer is a 500~nm piezoelectric ZnO thin film. (b) and (c) are strain profiles of the diamond devices simulated by COMSOL for a 3~GHz mode, with a 1 V$_{p}$ voltage driving source. The 20-$\mu$m and 10-$\mu$m-thick devices have a parabolic curved solid immersion lens with radius of curvature 40~$\mu$m and 20~$\mu$m respectively, providing both acoustic confinement and optical refraction suppression.}
\label{fig:1}
\end{figure}

As illustrated in Fig.~\ref{fig:1}(a), the device has a planar-convex structure. We design one side of the resonator with a curved surface, enabling it to confine three dimensional phonon modes with characteristic dimensions of 10~$\mu$m. The radius of curvature of the curved surface is twice the thickness of the substrate in a semi-confocal geometry, giving rise to the device name, SCHBAR. Compared to a planar cavity, the curved surface eliminates the requirement of boundary parallelism and, in principle, yields higher mechanical quality factors. Optically, the curved surface also acts as a solid immersion lens (SIL). It reduces substrate refraction and thus enhances light extraction from the defects inside the resonator\cite{hadden2010strongly}. The defects are microns below the surface and thus are well-protected from fabrication damage and surface effects. On the planar side of the device, we fabricate an integrated piezoelectric transducer and a microwave antenna that is aligned with the center of the resonator, allowing for acoustic and magnetic driving, respectively. The radius of the transducer has been designed to mode-match the waist of the confined acoustic wave, and the thickness of piezoelectric ZnO film is controlled to target a 3 GHz resonance mode which allows stable confinement.

We simulate the mechanical performance of the device using COMSOL. Fig.~\ref{fig:1}(b) and (c) show the strain profile of a $\sim$3~GHz acoustic mode for a 20~$\mu m$ and a 10~$\mu$m device, with a driving voltage amplitude at $V_{p}=1$~V. The simulation results show stable confinement of the acoustic wave and a peak strain around $5-10\times10^{-4}~\text{V}_{p}^{-1}$. 


\begin{figure}
\includegraphics{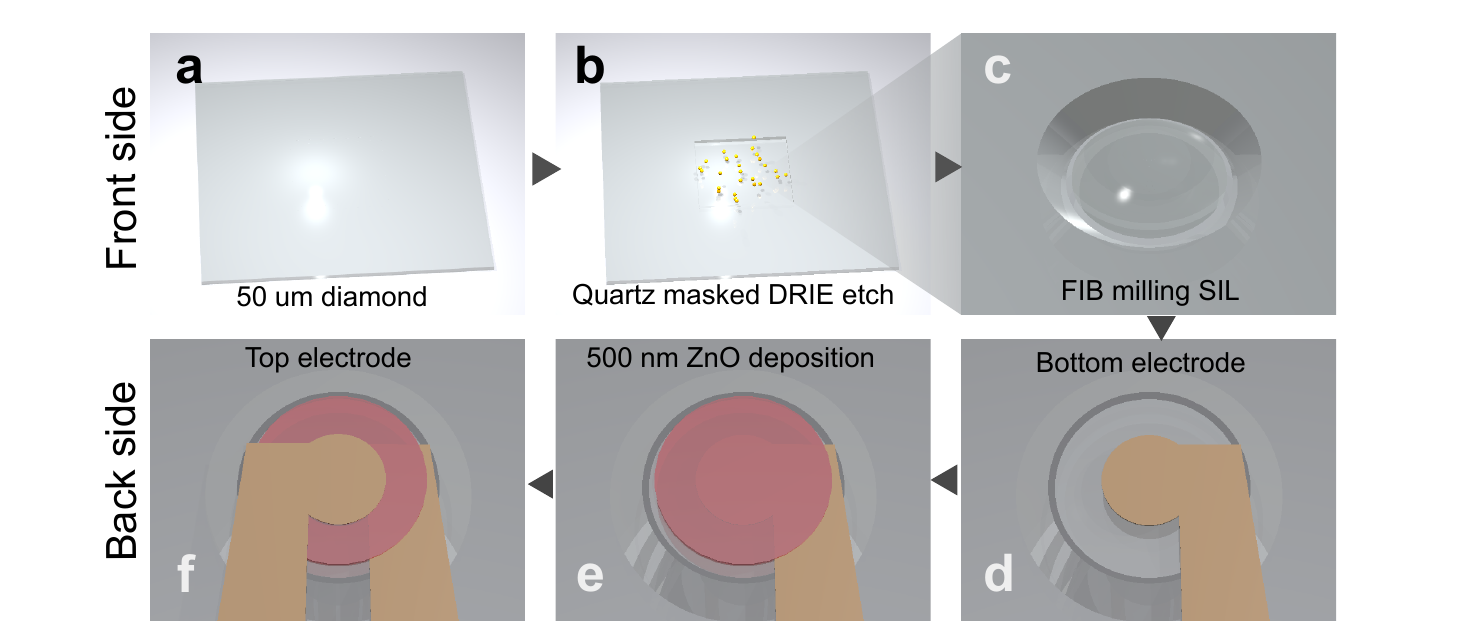}
\caption{Fabrication process flow of the device starting from (a) a 50-$\mu$m-thick double-side-polished diamond plate; (b) DRIE etch a diamond membrane down to 20~$\mu$m or 10~$\mu$m using Ar/Cl$_{2}$ and O$_{2}$ plasma, using a laser cut quartz mask; (c) Mill the parabolic SIL using focused Ga ion beam.  Post-milling plasma cleaning removes Ga damage and graphitization of substrate surface; (d-f) A ZnO piezoelectric transducer is fabricated on the backside of the SIL.}
\label{fig:2}
\end{figure}

We fabricated SCHBAR devices from both diamond and 4H-SiC substrates\cite{yu2018single}. For simplicity, here we restrict the process flow description to diamond, as presented in Fig.~\ref{fig:2}.  We start from a 50-$\mu$m-thick, double-side polished single crystal diamond plate (nitrogen concentration $<~1$~ppm). In the first step, we etch $5~\mu$m of diamond on each side of the substrate using Ar/$\text{Cl}_{2}$\cite{lee2008etching} and $\text{O}_{2}$ plasma\cite{friel2009control,appel2016fabrication, ruf2019optically} as a stress-relief etch to eliminate the residual polishing damage (see Supporting Information). A laser-cut quartz shadow mask is then used to mask the diamond for another 20- or 30-$\mu$m-deep etch on one side of the sample (etch rate $5~\mu$m/hr). After lifting the quartz mask, we end up with a 10~$\mu$m or 20~$\mu$m diamond membrane suspended in a 40~$\mu$m frame. Atomic force microscopy (AFM) shows a surface roughness of $<$0.3~nm. A focused gallium ion beam (FIB, 30~kV, 20~nA) is used to mill the parabolic SIL structure on the diamond membrane. After FIB milling, the diamond surface is substantially graphitized and contains implanted gallium atoms (20~nm in depth)\cite{bayn2011diamond}. We then etch away the top 100~nm of damaged diamond using Ar/$\text{Cl}_{2}$ plasma, followed by a 120~nm $\text{O}_{2}$ plasma etch to oxygen terminate the diamond surface. A boiling tri-acid bath containing equal parts of sulfuric, nitric and perchloric acid is used to further clean off any residual contamination on the diamond. Optical profilometry and laser-scanning confocal microscopy have been used to confirm the profile accuracy of the SIL. The diamond membrane is then flipped with the planar side facing up. A piezoelectric zinc oxide (ZnO) transducer, consisting of bottom electrode (10~nm/90~nm Ti/Pt), 500 nm ZnO, top electrode (10~nm/180~nm Ti/Pt), and a microwave antenna are then lithographically defined and sputtered to finish the device fabrication.

\begin{figure}
\includegraphics{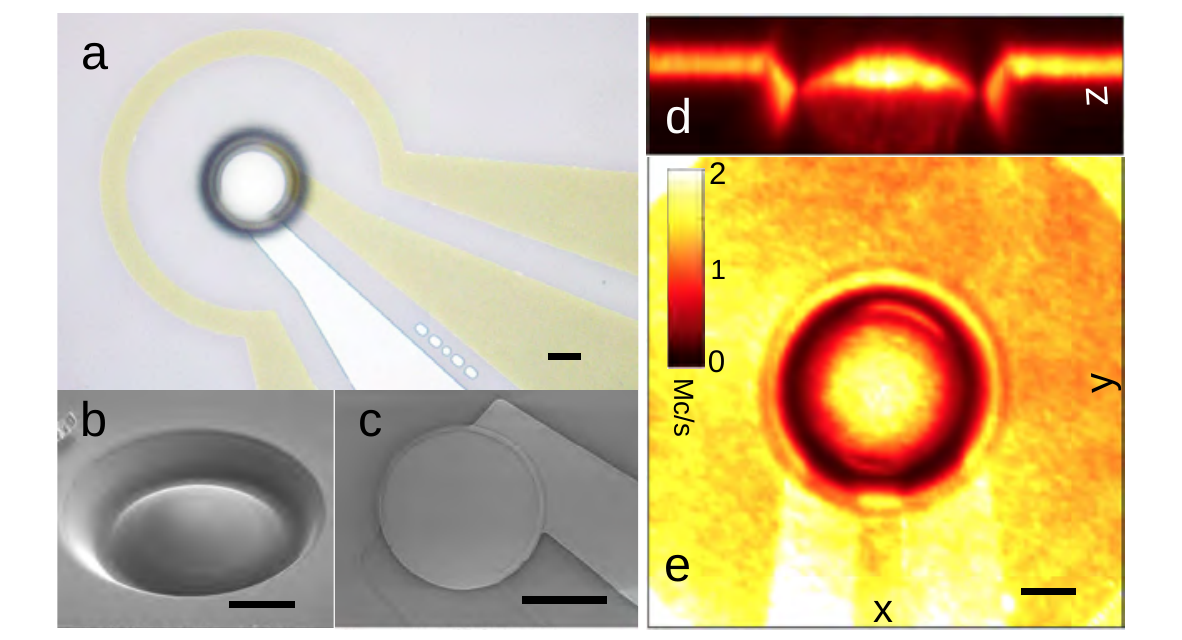}
\caption{Micrographs and photoluminescence images of the finished device on an optical grade diamond substrate. (a) Device viewed from SIL side. Encircling the devices is a microwave antenna used for magnetic resonance control of the spins inside the mechanical resonator. (b) SEM image of the milled solid immersion lens (radius of curvature 20~$\mu$m). (c) SEM image of the transducer on the backside of the SIL. (d) and (e) show the photoluminescence image in a cross section view and a front view of the device (10um thick), collected using a home-built confocal microscope. There is enhancement in fluorescence collection of the NV ensemble in the SCHBAR owing to the integrated SIL. The scale bars in all figures are 10 $\mu$m in length.}
\label{fig:3}
\end{figure}

Fig.~\ref{fig:3}(a-c) show the images of a finished diamond device. We measure the photoluminescence (PL) of the NV center ensemble inside the diamond SCHBAR using a home-built confocal microscope, where a 532~nm laser is used for excitation and a 630~nm long pass filter is used for PL collection in the phonon side band emission of NV centers. Fig.~\ref{fig:3}(d-e) show the cross section and the front PL scan of a 10 $\mu$m device at an incident laser power of 150 $\mu$W. There is a clear enhancement of PL collection inside the resonator. This, in combination with the spin measurement reported later in the letter, confirms that the NV centers are intact after fabrication.

\begin{figure}
\includegraphics{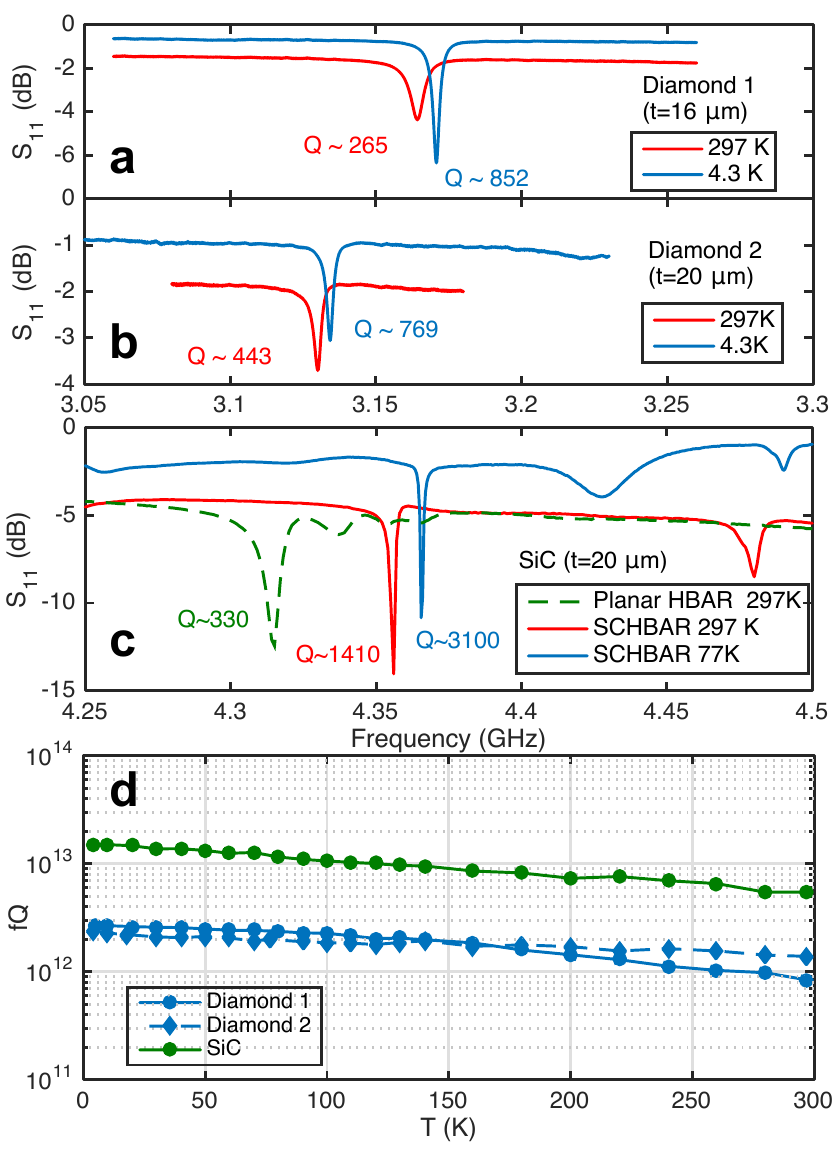}
\caption{(a-c) S-parameters characterization of the device using a vector network analyzer. Similar fabrication has been done for both (a-b) diamond and (c) 4H-SiC substrate. For the SiC device, the SIL is milled after the fabrication of transducer. Compared to a planar HBAR structure (green dashed line), the incorporation of a SIL increases the quality factor by more than three times.
(d) Quality factors are extracted from the measured S-parameters as a function of temperature.}
\label{fig:4}
\end{figure}

The device is then wire-bonded to a circuit board for electrical measurement. A vector network analyzer (VNA) is used to characterize the scattering parameter (S-parameter) of the device. When the resonator is driven on resonance, microwave power is dissipated in the resonator and converted into mechanical energy, launching acoustic waves and enabling the resonance to be detected as a decrease in the reflected microwave power. The electromechanical response of the device is well described by the modified Butterworth Van-Dyke model\cite{larson2000modified}, and the mechanical quality factor, $Q$, can be extracted from the VNA measurements. After mounting the devices in vacuum on a cold finger of a helium-flow cryostat, we perform electrical measurements as a function of temperature. The results are shown in Fig.~\ref{fig:4}. The frequency and quality factor product is $f\cdot Q>10^{12}$ for a 20-$\mu$m-thick diamond device at room temperature and $f\cdot Q>10^{13}$ for a 20-$\mu$m-thick SiC device at low temperature. For the SiC device, we change the order of fabrication to enable a test of the Q-enhancement due to the curved acoustic mirror. In this case, we mill the SIL after the piezoelectric transducer fabrication, enabling a separate measurement of both a planar and a semi-confocal version of the same device. We find a $>$3x improvement in the quality factor with the addition of the SIL. We have also found similar effects in 10-$\mu$m-thick diamond devices (see Supporting Information), indicating that the additional acoustic confinement adds in the suppression of acoustic diffraction loss. Theoretically\cite{zener1937internal,landau1937absorption,akhieser1939absorption,woodruff1961absorption,landau1986theory,lifshitz2000thermoelastic}, the quality factor of the current diamond device is limited by phonon scattering, $f\cdot Q$  $=3\sim4\times10^{13}$ at room temperature (see Supporting Information), which is an order of magnitude higher than the measured value. Possible sources for the extra damping can be surface dissipation and material losses within the piezoelectric transducer, which are not included in our estimate.

Next we characterize the defect spins in the resonator and their coupling to the phonon mode. Although we detected silicon vacancy (V$_{\text{Si}}$) center emission in the SiC device (see Supporting Information), indicating the possibility of acoustic control of defect system in SiC, here we focus on NV center spin measurements using a diamond SCHBAR.

At room temperature, both the ground and the excited states of an NV center are spin triplets. The ground-state electron spin $\ket{\pm1}$ state degeneracy can be lifted via their coupling to magnetic and strain fields (Fig.~\ref{fig:4}(a)), depicted by the Hamiltonian\cite{doherty2013nitrogen}:

 \begin{equation}
\label{eq:Hamfull}
\begin{split}
H_{\text{NV}}=&(D_{0}+d_{\parallel}\epsilon_{z})S^{2}_{z}+\gamma_{NV}\overrightarrow{B}\cdot\overrightarrow{S}+d_{\perp}[\epsilon_{y}(S_{x}S_{y}+S_{y}S_{x})-\epsilon_{x}(S^{2}_{x}-S^{2}_{y})],
\end{split}
\end{equation}
where $D_{0}=(2\pi)2.87$~GHz is the zero-field splitting, $\gamma_{\text{NV}}$=$(2\pi)2.8$~MHz/G is the gyromagnetic ratio, $\overrightarrow{S}$ is the electron spin of an NV center ($S=1$), $d_{\perp}(d_{\parallel})$ is the spin coupling strength to the transverse (longitudinal) strain field $\epsilon_{\perp}(\epsilon_{\parallel})$. Three sets of qubits can therefore be formed: magnetically-driven single quantum spin transitions between $\{\ket{0},\ket{-1}\}$ and $\{\ket{0},\ket{+1}\}$ states, and the mechanically-driven double quantum spin transition between $\{\ket{+1},\ket{-1}\}$ states\cite{macquarrie2013mechanical}. We use the latter one to quantitatively characterize the resonator performance and to study the defect electron-phonon coupling in a diamond SCHBAR device.

\begin{figure}
\includegraphics{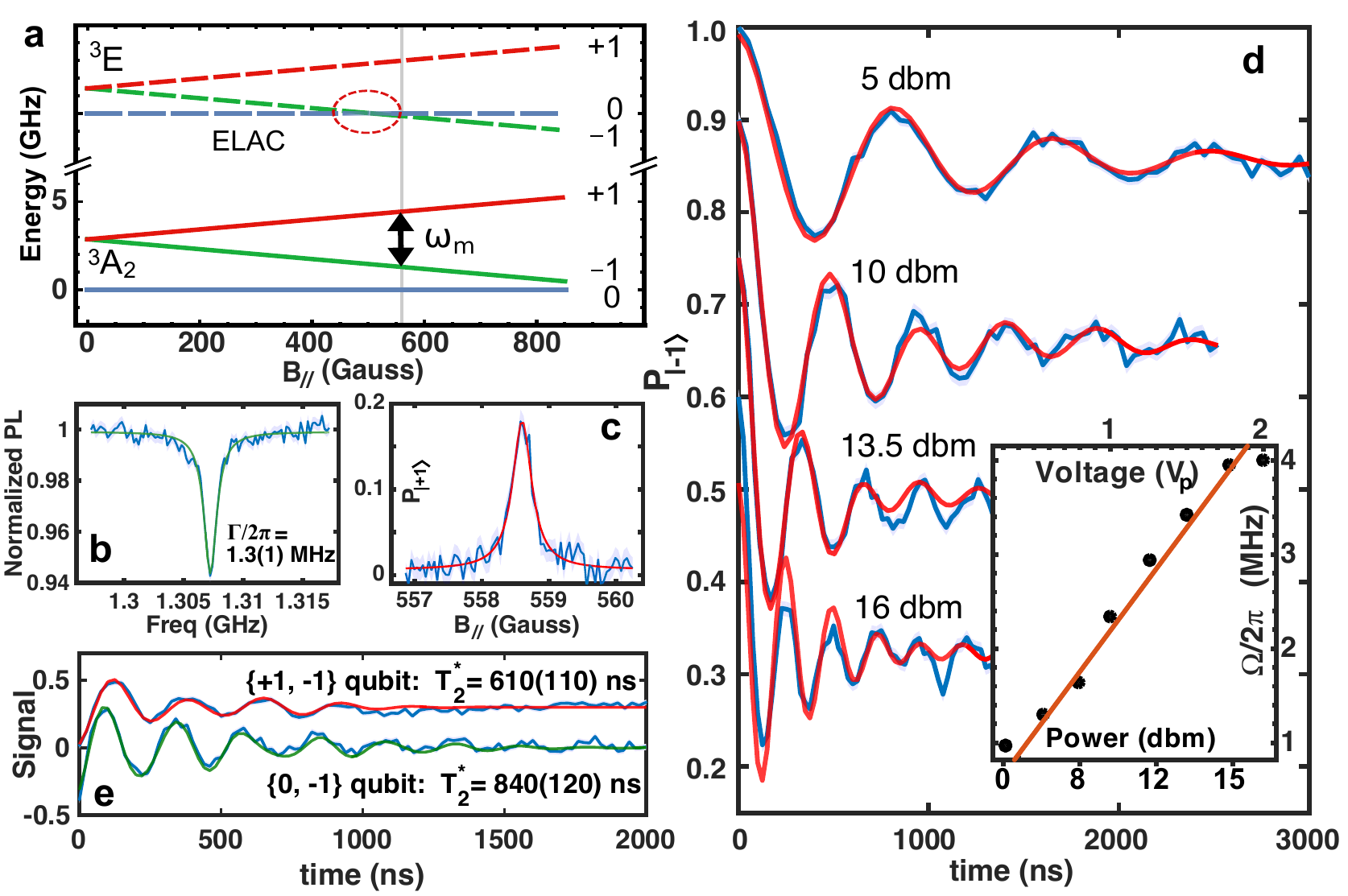}
\caption{(a) Energy level diagram of an NV center at room temperature. The targeted 3.13~GHz phonon mode is in resonance with the $\{\ket{+1}$, $\ket{-1}\}$ transition at magnetic fields around 558.6~G, close to the ELAC. (b) Optically detected magnetic resonance (ODMR) spectrum for the $\{\ket{0}$, $\ket{-1}\}$ transition reveals strong nuclear spin polarization of the NV ensemble. (c) ODMSR measurement also shows a single hyperfine peak, in agreement with (b). (d) Rabi oscillations between the $\{\ket{+1}$, $\ket{-1}\}$  states of the NV center ensemble induced by acoustic wave driving at 3.13~GHz taken at various driving powers. The inset is a linear fit to the Rabi frequency dependence on applied MW power to the transducer, giving a slope of $(2\pi)2.19(14)$~MHz/V$_{p}$. (e) Ramsey measurement of the $\{\ket{0}$, $\ket{-1}\}$ qubit state using magnetic driving, and $\{\ket{+1}$, $\ket{-1}\}$ qubit state using mechanical wave driving.}
\label{fig:5}
\end{figure}
All measurements are made at room temperature using a 20-$\mu$m-thick diamond SCHBAR. We first apply an external magnetic field of 558.6~G, axially aligned to the N-V axis. This splits the $\ket{+1}$ and $\ket{-1}$ ground spin states of the NV center ensemble due to the Zeeman effect~(Fig.~\ref{fig:4}(a)). At 558.6~G, the spin $\ket{+1}$ and $\ket{-1}$ state separation is in resonance with the phonon mode frequency, $\omega_{\text{m}}$. We then launch acoustic waves in the resonator by applying microwave power to the transducer at the resonator frequency 3.13~GHz. Coherent spin transitions can thus be induced by the acoustic wave through the transverse spin-strain coupling\cite{macquarrie2013mechanical,macquarrie2015coherent}.

Note that the operating magnetic field strength is close to the excited-state level anti-crossing (ELAC) at 508~G\cite{jacques2009dynamic}. By optical pumping in a well-aligned magnetic field, we can induce a strong polarization of the $I=1$ nuclear spin of $^{14}$N into its $\ket{m_{I}=+1}$ state (Fig.~\ref{fig:4}(b)). By working with only a single hyperfine state, we greatly improve the signal-to-noise ratio and reduce the complexity of the system dynamics.

Now we test the mechanical driving efficiency and coherence. To pinpoint the mechanically-driven spin resonance, we perform optically detected mechanical spin resonance (ODMSR) measurements as described in Ref.~\citen{macquarrie2013mechanical} (see Supporting Information). We observe a single resonance peak in the spectrum, confirming the strong nuclear spin polarization within the NV center ensemble. We then perform mechanically-driven Rabi measurements of the $\ket{+1}$ and $\ket{-1}$ states as a function of microwave power applied to the transducer. We observe high contrast Rabi oscillations at a rate $(2\pi)2.19(14)$~MHz/V$_{p}$, where V$_{p}$ is the peak voltage applied to the input port of our device. Transverse NV center spin-strain coupling, $d_{\perp}$, has been previously measured at around 20~GHz/strain\cite{ovartchaiyapong2014dynamic,teissier2014strain}, enabling us to estimate that the current device has a high power-to-strain efficiency of $ 1.59(14)\times10^{-4}~\text{V}_{p}^{-1}$ (see Supporting Information). This value is close to the COMSOL simulation result, and compared to an HBAR device\cite{macquarrie2015coherent}, the power-to-strain efficiency has improved by 60-160x. Lastly, we measure the spin coherence within the mechanically-controlled $\{\ket{+1}$, $\ket{-1}\}$ subspace and within the magnetically-controlled $\{\ket{0}$, $\ket{-1}\}$ subspace (Fig.~\ref{fig:5}(e)). We find $T^{*}_{2}$ is 610(110)~ns and 840(120)~ns, respectively.


With the demonstrated voltage-to-strain transduction, the diamond SCHBAR device has roughly the same efficiency for spin-strain driving as the microwave antenna has for magnetic driving. This is promising for using mechanical driving as an added resource, for example, strong mechanical driving of NV center spins can be used to extend the coherence time via continuous dynamical decoupling\cite{macquarrie2015continuous,barfuss2015strong}. Additionally, the ability to perform pulsed spin control protocols within the $\{\ket{+1}$, $\ket{-1}\}$ subspace, which are enabled by sizable Rabi frequencies and $\sim$40~ns ring-up/down times, makes quantum control a strong application of our device. These capabilities will be a useful resource for magnetic sensing in the double-quantum basis\cite{taylor2008high} and for non-magnetic sensing modalities such as inertial sensing\cite{ajoy2012stable}. Apart from fast mechanical control of defect spins, a SCHBAR operates at a high enough frequency to be cooled to its mechanical ground state using a dilution refrigerator ($\sim$~100~mK). In combination with spin ensemble enhancement to phonon coupling, the device can potentially be used for studying spin-phonon coupling at a single phonon level\cite{blais2004cavity} (see Supporting Information). Improvements to the device quality factors and power handling are possible by replacing ZnO with AlN. Further enhancement in single spin-phonon coupling can be obtained by further reduction of the device volume, however, it must accompany a shift to a higher resonator frequency. An alternative approach is to incorporate color centers with intrinsically stronger spin-phonon coupling, such as the diamond SiV center\cite{meesala2018strain}.

In conclusion, we report the design, fabrication and testing of a diamond (SiC) SCHBAR device. SCHBAR marks the first step towards a micro-scale BAW resonator device. We obtain $f\times Q>10^{12}(10^{13})$ for a diamond (SiC) SCHBAR and demonstrate efficient and high-rate phonon-induced NV center Rabi oscillations at $(2\pi)2.19(14)$~MHz/V$_{p}$. The device enables direct circuit integration and it is ideal for fast mechanical control of defect spins. Further integration of a dense spin ensemble along with improvements in the device quality factor could enable study of defect spin-phonon interaction at the quantum level.

\begin{acknowledgement}

Research support was provided by the Office of Naval Research (Grant N000141712290) and by the DARPA DRINQS program (Cooperative Agreement $\#$D18AC00024). Any opinions, findings, and conclusions or recommendations expressed in this publication are those of the author(s) and do not necessarily reflect the views of DARPA. R.S.D. acknowledges support of the Air Force Office of Scientific Research Hybrid Materials MURI under award number FA9550-18-1-0480. Device fabrication was performed in part at the Cornell NanoScale Science and Technology Facility, a member of the National Nanotechnology Coordinated Infrastructure, which is supported by the National Science Foundation (Grant ECCS-15420819), and at the Cornell Center for Materials Research Shared Facilities which are supported through the NSF MRSEC program (Grant DMR-1719875). 

\end{acknowledgement}


\newpage
\begin{center}
\section{Supporting Information
  \\ 
\begin{large} 
  Engineering electron-phonon coupling of quantum defects to a semi-confocal acoustic resonator
\end{large} }
\end{center}



\section{SCHBAR VS HBAR}
\begin{figure}
\includegraphics{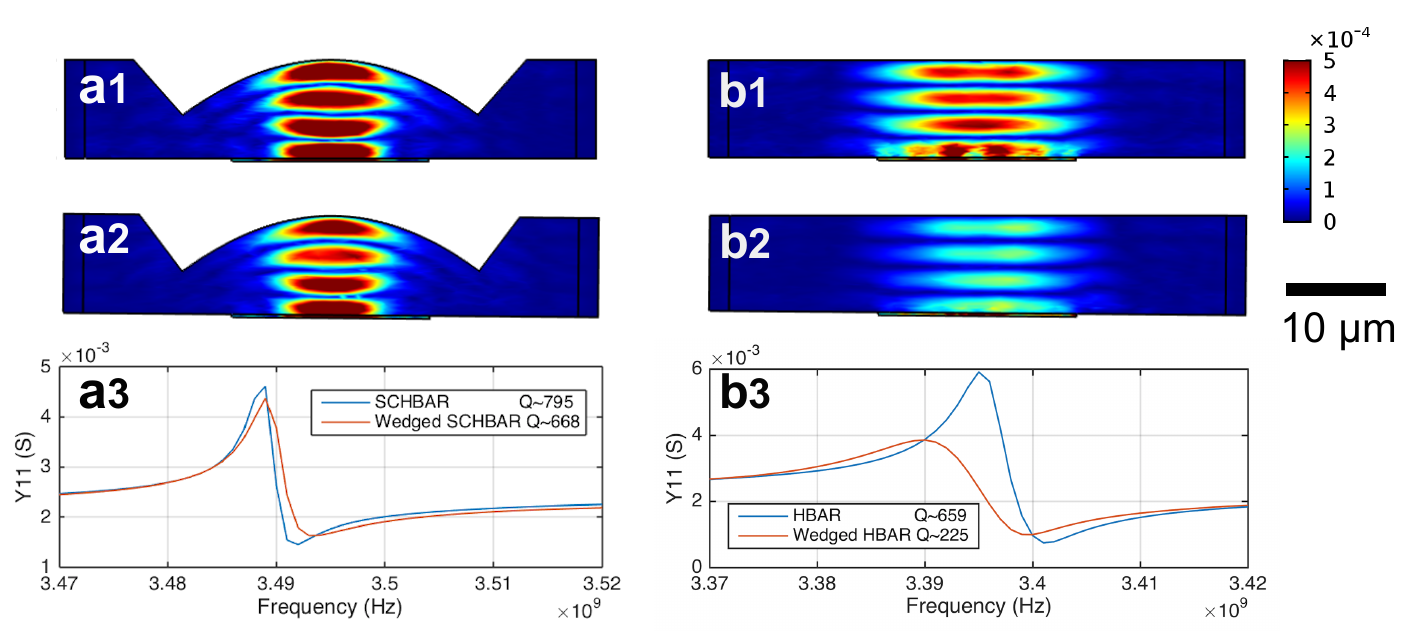}
\caption{(a) and (b) are COMSOL simulation (a tetrahedral finite element mesh is used, with 0.5~$\mu$m element size in the vertical direction) for a SCHBAR and an HBAR of thickness 10 $\mu m$, respectively. In (a2) and (b2), a 0.46 degree tilt angle is introduced in one surface of the substrate. (a3) and (b3) are the simulated admittance curves. The SCHBAR shows immunity to a wedged substrate as compared to a planar HBAR.}
\label{fig:COMSOL}
\end{figure}

We compare the performance of an SCHBAR to an HBAR using COMSOL modeling. Fig.~\ref{fig:COMSOL} shows the simulation results of a 10-$\mu m$-thick planar HBAR and a 10-$\mu m$-thick SCHBAR device with a 1~V$_{p}$ drive, where V$_{p}$ is peak voltage. In comparison to the planar HBAR, the SCHBAR has stronger phonon mode confinement and a larger strain amplitude. An important factor that determines the performance of an HBAR device is the surface parallelism. A slight thickness variation across the substrate can lead to significant phonon leakage and a reduced quality factor. In Fig.~\ref{fig:COMSOL}(a-b2), we introduce a 0.46 degree tilt angle (0.4~$\mu$m thickness variation across 50~$\mu$m) to one surface of the substrate. In contrast to the drastic decrease of quality factor in an HBAR device, there is little impact on the SCHBAR $Q$-factor from the substrate thickness variation. This is beneficial for diamond devices because it alleviates the stringent requirement for alignment in substrate polishing which helps to achieve a high quality factor.

\section{Optical collection efficiency enhancement}

\begin{figure}
\includegraphics{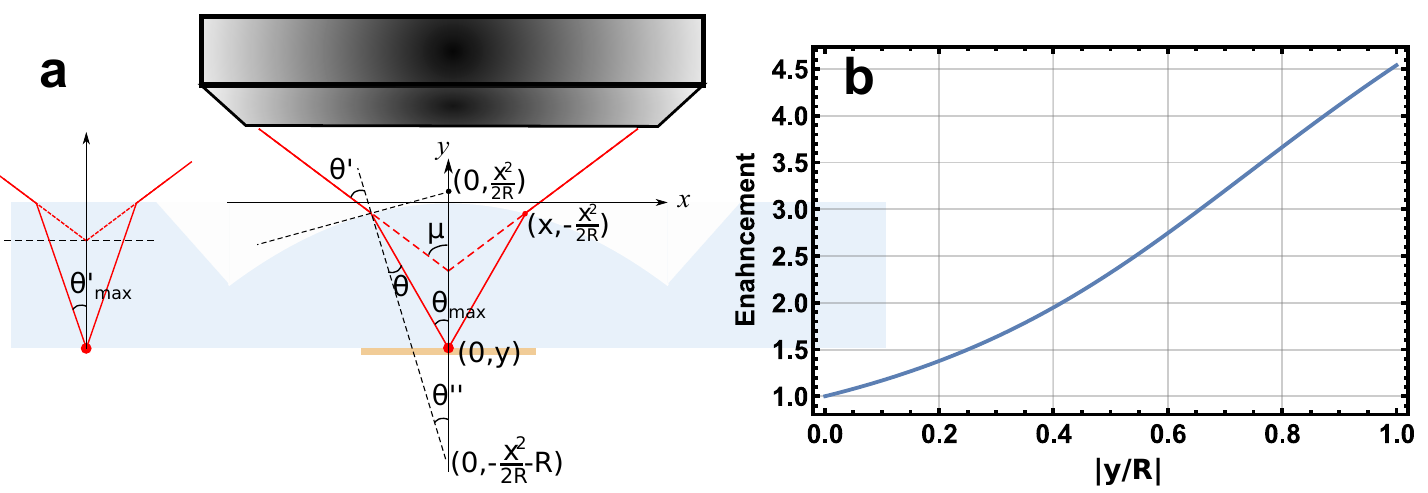}
\caption{(a) Ray optics schematics of the photon collection. (b) Enhancement in optical collection efficiency of a parabolic SIL compared to a planar substrate as a function of depth. NA=0.8 is assumed for the objective.}
\label{fig:SIL}
\end{figure}

Using a ray optics calculation, we numerically solve the collection solid angle as a function of depth in the device. At the critical angle, the exit ray angle $\theta+\theta^{\prime}$ matches the numerical aperture of the lens, $\mu=\sin^{-1}\text{NA}$. The schematics is shown in Fig.~\ref{fig:SIL}(a), with the following constraints applied:
\begin{equation}
\label{eq:SIL}
\begin{split}
\mu&=\theta^{\prime\prime}+\theta^{\prime}
\\
\theta^{\prime\prime}&=\tan^{-1}\frac{x}{R}
\\
n&=\sin\theta^{\prime}/\sin\theta
\\
\theta&=\tan^{-1}(x/(\frac{-x^{2}}{2R}-y))-\theta^{\prime\prime}
\end{split}
\end{equation}
where $R$ is the radius of curvature of the SIL at $x=0$, $n$ is the substrate refractive index, $y$ is the depth inside the SIL.

The solution of $\theta^{\prime\prime}$ can be numerically solved, and the maximum collection angle is $\theta_{\text{max}}=\tan^{-1}(\frac{\tan \theta^{\prime\prime}}{-\frac{1}{2}\tan^{2}\theta^{\prime\prime}-y/R})$. Compared to a planar structure, $\theta^{\prime}_{\text{max}}=\sin^{-1}(\sin(\text{NA})/n)$, the enhancement in fluorescence collection is shown in Fig.~\ref{fig:SIL}(b), evaluated as $\frac{1-\cos(\theta_{\text{max}})}{1-\cos(\theta^{\prime}_{\text{max}})}$.

\section{Diamond etching and characterization}
We use Ar/Cl$_{2}$\cite{lee2008etching} combined with O$_{2}$ plasma to etch diamond. Mechanically polished diamond surfaces have micro-scale defects (particles and pits) and polishing streaks from the lapping process. Without proper treatment, O$_{2}$ plasma tends to etch anisotropically around the defects, leaving either a grass-like rough surface or pitted surface with craters\cite{friel2009control}. Ar/Cl$_{2}$ etches diamond isotropically and can effectively remove the surface contaminants, smoothening the diamond surface\cite{ruf2019optically}. To compensate for the slow etch rate of Ar/Cl$_{2}$ plasma, we switch to O$_{2}$ plasma after Ar/Cl$_{2}$ plasma etch and alternate to Ar/Cl$_{2}$ plasma intermittently for a short time to remove possible surface contamination\cite{appel2016fabrication} from, for instance, sputtered particles in the etcher chamber. Detailed operating parameters for diamond etch are listed in Table.~\ref{table:etch1}-\ref{table:etch2}.

\begin{table}
\begin{tabular}{lllllll}
\hline
{Gas} & Flow rate & Pressure & ICP/RIE & DC bias & etch rate & {Selectivity to SiO$_{2}$} \\
                     & (sccm)    & (mTorr)  & (W)     & (V)     & (nm/min)  &                                           \\ \hline
Ar/Cl$_{2}$          & 25/40     & 8        & 600/100 & 217     & 34(1)     & 0.72                                      \\ 
O$_{2}$              & 60        & 10       & 950/50  & 132     & 118(2)    & $\infty$                                      \\ \hline
\end{tabular}
\caption{Diamond plasma etch parameter}
\label{table:etch1}
\end{table}

\begin{table}
\begin{tabular}{lll}
\hline
Etch Type                      & Process                                             & Etch rate/depth (include plasma off time) \\ \hline
{Stress relief} & 35 min Ar/Cl$_{2}$+20 min O$_{2}$ & {4.9 $\mu$m}        \\ 
                               & +5 min Ar/Cl$_{2}$+10 min O$_{2}$                   &                                     \\ 
DRIE                           & 5 min Ar/Cl$_{2}$+20 min O$_{2}$                    & 5~$\mu$m/hr                         \\ 
Post FIB clean                &  3 min Ar/Cl$_{2}$+1 min O$_{2}$                                                   & 220~nm          \\\hline
\end{tabular}
\caption{Diamond etch process}
\label{table:etch2}
\end{table}

\begin{figure}
\includegraphics{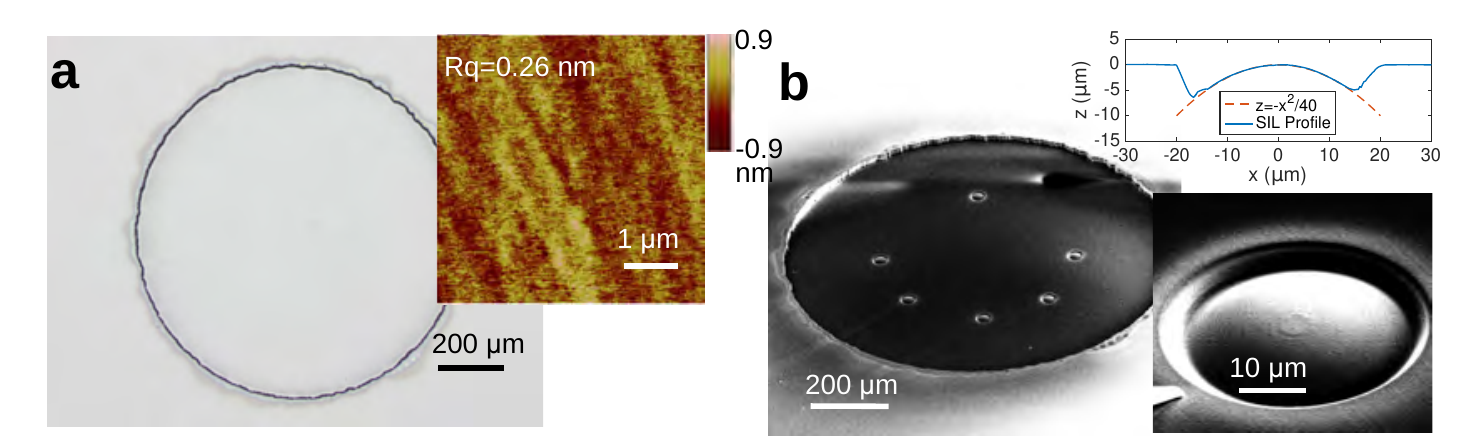}
\caption{(a) Optical image of a circular defect-free diamond membrane. The inset shows an AFM scan of the surface. (b) SEM images of SILs milled onto the membrane using focused ion beam. The inset shows the SIL profile measured by a confocal microscope.}
\label{fig:etch}
\end{figure}

After quartz masked diamond etch, we perform optical profilometry (Zygo NewView 7300) to measure the etch depth and use a laser scanning confocal microscope (Keyence VK-X260) for membrane thickness characterization. After etching, AFM measurement shows that the surface roughness is less than 0.3~nm (Fig.~\ref{fig:etch}).

\section{Quality factor limit analysis}
In general, the quality factor of a mechanical resonator can be determined from various sources: 1) air damping, 2) clamping loss, 3) thermoelastic dissipation (TED), 4) phonon scattering, 5) lattice friction from crystalline defect, etc. While air damping is negligible in a sufficiently high-frequency resonator and can be eliminated by operating in vacuum, the rest require careful design and engineering to minimize. Clamping loss arises when the resonator anchor is stressed and can thus be avoided by designing the resonator to have a stress-free boundary, such as a tuning fork. In our SCHBAR, the boundary is close to a stress free condition from the SIL confinement. TED occurs due to the strain-gradient-induced heat flow inside resonator. A few approaches have been demonstrated to address TED: 1) design a resonator with isotropic strain or a pure shear mode resonator that has no volume change, 2) operate in an environment where the thermal expansion coefficient of the resonator material is zero, 3) block heat flow by cutting slots in a beam resonator, 4) operate resonator at a frequency away from the thermalization rate, $1/\tau_{th}$\cite{lifshitz2000thermoelastic}. For low-frequency beam resonators, TED can be a major limiting factor to the quality factor at 100~K and above. Lifshitz\cite{lifshitz2000thermoelastic} derived the exact solution of TED based on Zener's work\cite{zener1937internal},

\begin{equation}
\label{eq:Zener}
\begin{split}
fQ_{\text{Beam}}&=\frac{fC_{v}}{E\alpha^{2} T}\left(\frac{6}{\xi^{2}}-\frac{6}{\xi^{3}}\frac{\sinh\xi+\sin\xi}{\cosh\xi+\cos\xi}\right)^{-1}
\end{split}
\end{equation}
where $C_{v}=\rho C_{sp}$ is the heat capacity per unit volume, $C_{sp}$ is the specific heat, $E$ is the Young's modulus of the material, $\alpha$ is the thermal expansion coefficient at temperature $T$. $\xi=\pi\sqrt{\omega_{m}\tau_{th}/2}$, $\tau_{th}=b^{2}C_{v}/(\pi^{2}\kappa)$ is the thermal relaxation time, $b$ is the width of the beam, $\kappa$ is the thermal conductivity.

For a high-frequency (here 3~GHz) bulk acoustic wave resonator (equivalently an infinitely thick beam resonator), acoustic attenuation caused by TED is significantly less because of a longer thermalization time, $\tau_{th}$, exceeded by resonator frequency (applies for $T>10$~K), $\omega\tau_{th}>1$.
The theoretical limit of Q is derived for a longitudinal wave in an isotropic media by Landau\cite{landau1986theory},

\begin{equation}
\label{eq:Landau}
fQ_{\text{BAW}}=(1-\frac{4c_{t}^{2}}{3c^{2}_{l}})^{-2}\frac{C_{v}^{2}}{2\pi\kappa\rho\alpha^{2}T},
\end{equation}
where $c_{t}$ and $c_{l}$ are shear and longitudinal speed of sound.

\begin{figure}
\includegraphics{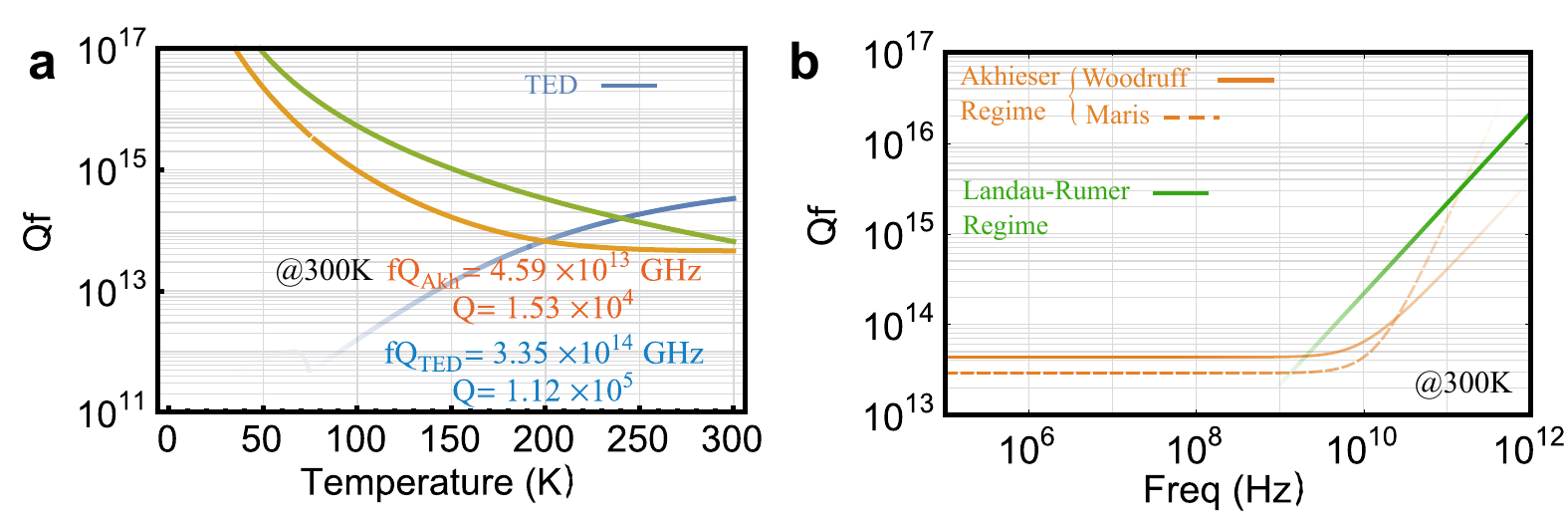}
\caption{(a) Theoretical calculation of $fQ$ limit of a 3~GHz diamond BAW resonator based on experimental data. TED limit calculation applies $>$150~K, where $\tau_{th}>\tau_{p}$. (b) Phonon interaction limited $fQ$ calculation for diamond at room temperature.}
\label{fig:Q}
\end{figure}

While the TED limit for a bulk acoustic wave device is high, $f\times Q\sim3\times10^{14}$ for diamond at 300~K (Fig.\ref{fig:Q}(a)), phonon scattering process enforces another fundamental limit to the quality factor, theorized by Akhiezer\cite{akhieser1939absorption}, Landau and Rumer\cite{landau1937absorption,maris1971interaction}:

\begin{equation}
\label{eq:Akhiezer}
\begin{split}
fQ_{\text{AKE}}&=\left\{\begin{matrix}
\frac{\rho c^{2}}{C_{v}\gamma^{2}T}\frac{3}{4\pi\tau_{p}}\frac{2\omega\tau_{p}}{\tan^{-1}2\omega\tau_{p}}~(\text{Woodruff\cite{woodruff1961absorption}})\\ 
\frac{\rho c^{2}}{C_{v}\gamma^{2}T}\frac{1+\omega^{2}\tau_{p}^{2}}{2\pi\tau_{p}}~(\text{Maris\cite{maris1971interaction}})
\end{matrix}\right.,~~~~\hbar\omega<k_{B}T,~\omega\tau_{p}<1
\\
fQ_{\text{L-R}}&=\frac{15}{8\pi^{7}}\frac{\rho c^{5}h^{3}}{\gamma^{2}}\frac{\omega}{(k_{B}T)^{4}},(\text{longitudinal wave}\cite{maris1971interaction})~~~~~~~~\hbar\omega<k_{B}T,~\omega\tau_{p}>1
\end{split}
\end{equation}
where $\gamma$ is the Gr$\ddot{\text{u}}$neisen's parameter, $\tau_{p}=3\kappa/C_{v}c^{2}$ is the phonon relaxation time. $h$ is Plank constant, $k_{B}$ is Boltzmann constant. Note that Woodruff's result is 1.5 times higher than Maris's at low frequency, and transits to the same frequency scaling as Landau-Rumer's result for high frequency regime (Fig.~\ref{fig:Q}(b)).

For a 3~GHz diamond SCHBAR, $\omega\tau_{p}\sim0.2$ at room temperature, close to the transition from Akhiezer regime to Landau-Rumer regime, as shown in Fig.~\ref{fig:Q}(b). The quality factor limit is set by phonon-phonon interactions, $f\times Q=3-4\times10^{13}$ at room temperature, an order of magnitude higher than our measurement in the fabricated diamond SCHBAR. We attribute possible quality factor limiting sources to transducer loss and surface dissipation. Compared to an HBAR, SCHBAR devices have increased surface-to-volume ratio and less fraction of energy stored inside the substrate. While diamond etching has been optimized to achieve a smooth surface (roughness $<$0.3~nm), the ZnO film in the transducer is comparably rough (roughness$\sim$7~nm). Metal electrodes on the transducer and grains in ZnO can also cause significant phonon damping.

\section{10~$\mu$m device and further device scale-down }

\begin{figure}
\includegraphics{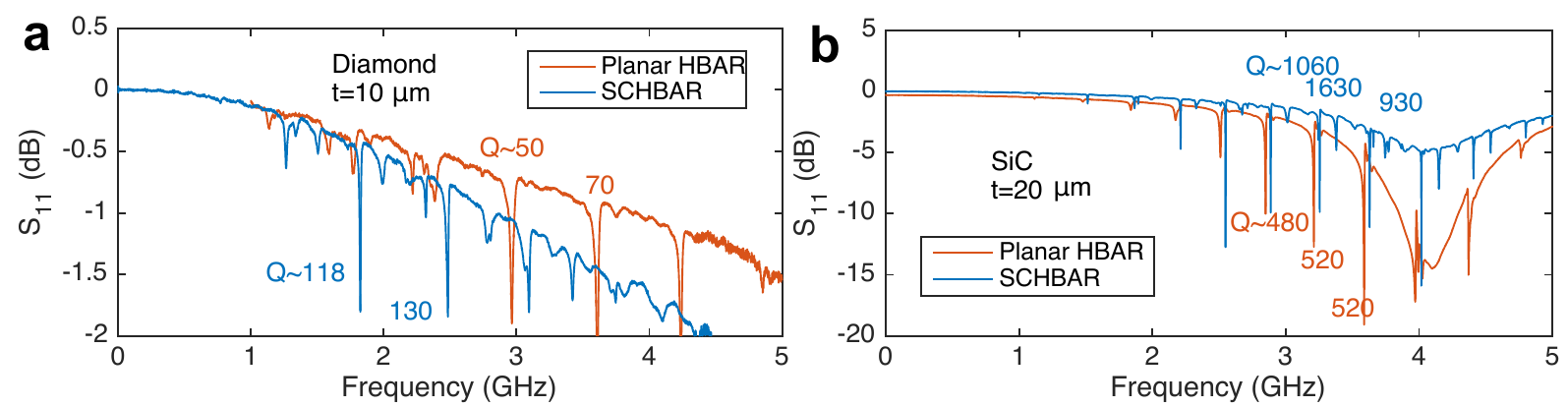}
\caption{Room temperature S-parameter measurements for (a) a diamond SCHBAR and (b) a SiC SCHBAR.}
\label{fig:10um}
\end{figure}

We fabricated diamond SCHBAR devices on a $t$=10~$\mu$m thick diamond membrane. Similar to the process that we used for the SiC SCHBAR, we mill the SILs after backside transducer fabrication. The S-parameter measurement results are shown in Fig.~\ref{fig:10um}. Compared to a planar HBAR of the same thickness, the SIL does enhance certain modes ($\sim$2~GHz) in the 10~$\mu$m diamond SCHBAR. However, the quality factors, $Q=100-200$, are overall lower than we found in the 20-$\mu$m-thick SCHBAR device. We observe that the SIL enhancement is more obvious in a 20~$\mu$m SiC SCHBAR process ($t/\lambda>4$), suggesting the semi-confocal geometry works better for a thicker substrate at the same frequency range. This also suggests that to obtain a high-performance SCHBAR at a smaller scale ($<10~\mu$m), a SCHBAR needs to operate at higher frequency ($t/\lambda>2,~f>2c/t$).

\section{V$_{\text{Si}}$ emission spectra from SiC SCHBAR}

\begin{figure}
\includegraphics{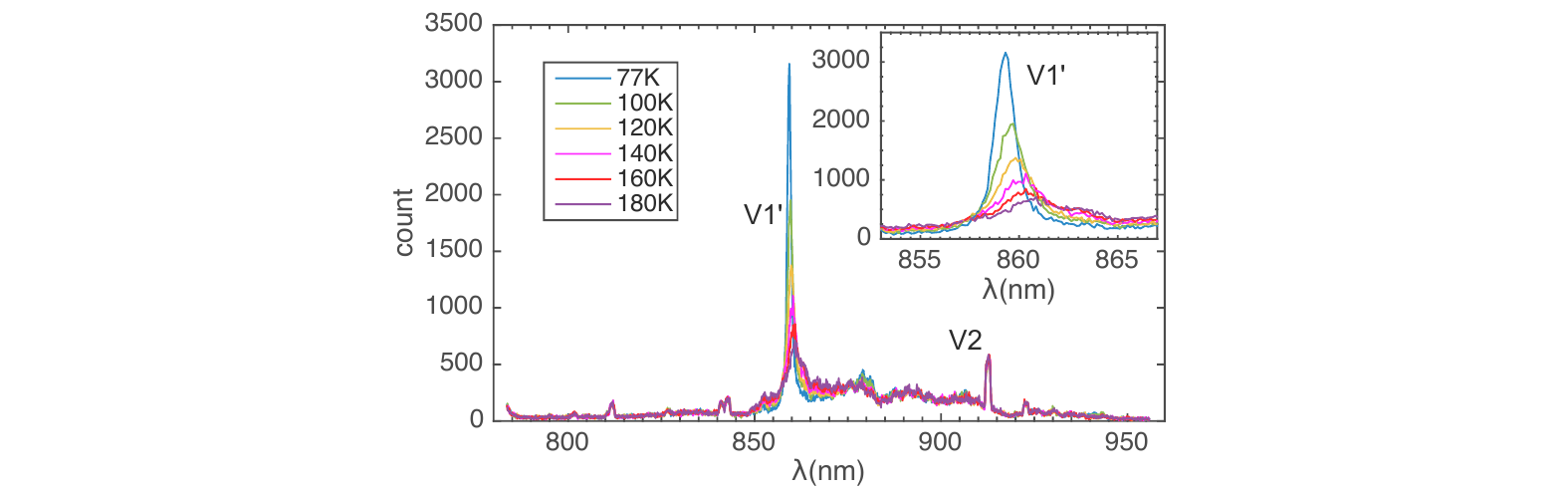}
\caption{Emission spectra of V$_{\text{Si}}$ center ensemble in SiC SCHBAR device taken at various temperature. The 860~nm and 917~nm peak are zero-phonon lines of V1' and V2 type V$_{\text{Si}}$ center.}
\label{fig:VSi}
\end{figure}

Similar to diamond, 4H-SiC substrate has high quality mechanical property and is a host to divacancy (V-V) centers and silicon vacancy (V$_{\text{Si}}$) centers, both of which contain electron spin with long coherence time\cite{christle2015isolated,widmann2015coherent}. We experimentally probe the defect emission spectra of the SiC SCHBAR at the near infrared range (800-950~nm) where zero-phonon lines of V$_{\text{Si}}$ centers are located. In a home-built confocal microscope setup, we off-resonantly excite the defects with a 780~nm pump laser at 0.6~mW. The emission of the V$_{\text{Si}}$ centers is filtered by a long pass filter (cutoff wavelength at 800~nm) and then analysed with a spectrometer. We identify both V1 and V2 type of V$_{\text{Si}}$ centers in the recorded spectra (Fig.~\ref{fig:VSi}).

\section{Pulse sequence for mechanical spin driving}

\begin{figure}
\includegraphics{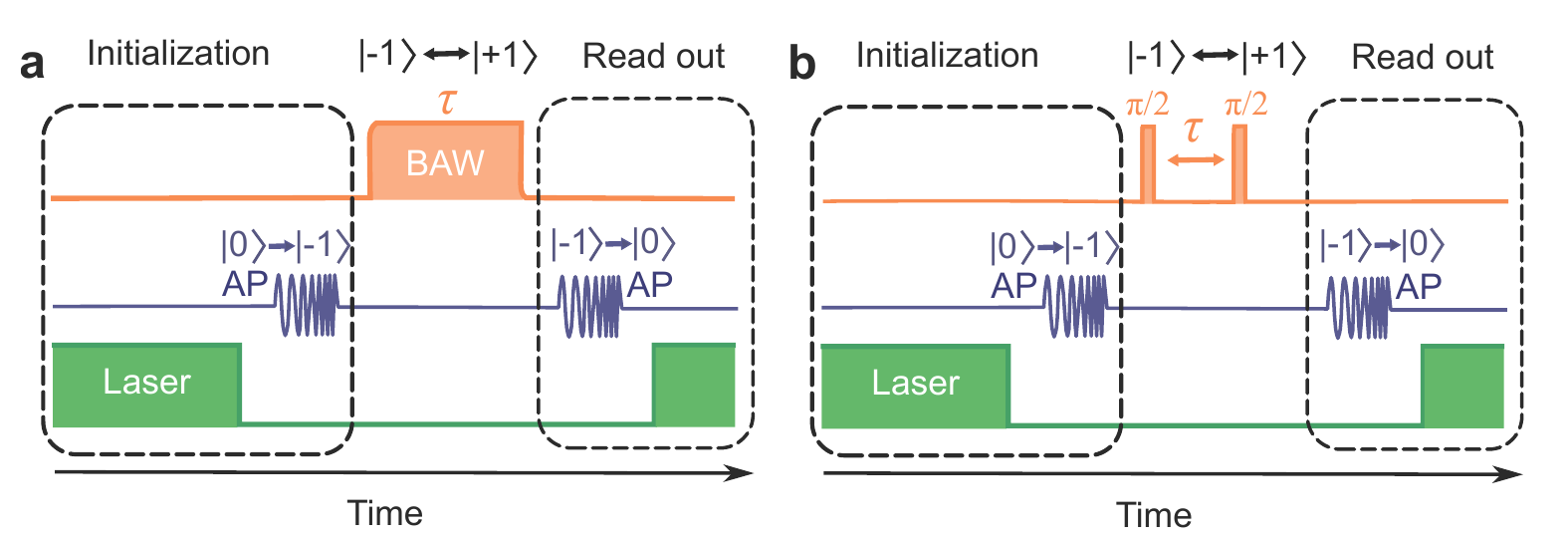}
\caption{(a) Pulse sequence used for ODMSR measurement and mechanically-driven spin Rabi oscillation. The orange, blue and green color represent acoustic wave generated by the piezoelectric transducer, magnetic control field from microwave antenna and optical pulse from the 532~nm laser, respectively. (b) Pulse sequence for mechanical Ramsey coherence measurement.}
\label{fig:sequence}
\end{figure}

The implemented mechanical spin driving scheme is illustrated in Fig.~\ref{fig:sequence}(a). We first initialize the NV centers into $\ket{m_{s}=0}$ state via optical polarization. Next, we use an adiabatic passage (AP) pulse to shelve the spins into $\ket{m_{s}=-1}$ state. We pulse on the acoustic wave for a period of time, $\tau$, to induce spin population transfer between $\{\ket{m_{s}=-1}$, $\ket{m_{s}=+1}\}$ states. After that, another AP pulse shuttles the residual $\ket{m_{s}=-1}$ spin population back to $\ket{m_{s}=0}$ state, which is then read out optically. The sequence is repeated 80,000x for a single measurement. In ODMSR experiment, we fix $\tau\sim$ 3~$\mu$s, microwave power applied to transducer at 0~dbm, and take measurements as a function of external magnetic field. In mechanically-driven Rabi oscillation experiments, we repeat the sequence and measure time evolution of the spin population as we vary the pulse duration $\tau$ of the acoustic wave. The mechanical Ramsey coherence measurement is implemented in a similar way (Fig.~\ref{fig:sequence}(b)), where two $\pi/2$ mechanical pulses with varying time delay, $\tau$, are used. For the measurement in the main text, $\tau_{\frac{\pi}{2}}$=197~ns. We further introduce a phase shift to the second $\pi/2$-pulse by $\omega_{rot}(\tau + \tau_{\frac{\pi}{2}})$, where $\omega_{rot}$ is the equatorial rotating frequency of the spin in the Bloch sphere. This extra phase shift introduces a known periodicity to the measurement that aids visualization of the decay envelope. Similar protocols have been described in ref.\citen{macquarrie2015coherent}.

The contrast of the recorded $\{\ket{m_{s}=+1}$, $\ket{m_{s}=-1}\}$ qubit Rabi oscillation is around 40\% of that for a magnetically driven $\{\ket{m_{s}=0}$,$\ket{m_{s}=-1}\}$ qubit. We attribute this to the fact that the acoustic standing wave inside the resonator is spatially inhomogeneous, with a periodical anti-node every half wavelength, $\sim2.5~\mu$m. The depth resolution of our objective is $\sim 3.5~\mu$m inside diamond. The resolution is improved deep inside the SIL to be $\sim 2.2~\mu$m. In the acoustic spin driving measurement, the probed volume includes both high strain (anti-node) and un-strained (node) area. Considering that the acoustically-driven Rabi oscillation rate and the contrast scales proportionally with the strain amplitude, the acquired signal is an ensemble average over the Rabi oscillations across both a node and an anti-node of the acoustic mode, leading to a reduced contrast in measurement.

\section{Strain evaluation}
In the fabricated diamond SCHBAR device, stress and strain ($\epsilon$) generated from the transducer are primarily perpendicular to the (001) cut face of the diamond crystal. We define the crystal basis as X: [-110] Y:[001] Z:[110]. After taking into account the Poisson ratio, $\nu\simeq0.1$, the strain tensor can be written as,
\begin{equation}
\label{eq:strain}
\mathbf{E}= \begin{pmatrix}
-\nu\epsilon & 0 & 0\\ 
 0&  \epsilon&0 \\ 
 0& 0 & -\nu\epsilon
\end{pmatrix}
\end{equation}

Transforming into N-V axis, for example, $x=(-\sqrt{2}/2, 1, 0), y = (0, 0, 1), z= (\sqrt{2}, 1, 0)$, we have
\begin{equation}
\label{eq:strainNV}
\mathbf{E}^{\prime}=  \begin{pmatrix}
\frac{2-\nu}{3}\epsilon & 0 & \frac{\sqrt{2}(1+\nu)}{3}\epsilon\\ 
 0&  -\nu\epsilon&0 \\ 
 \frac{\sqrt{2}(1+\nu)}{3}\epsilon& 0 & \frac{1-2\nu}{3}\epsilon
\end{pmatrix}.
\end{equation}
The effective coupling to an NV center spin is then
$\Omega_{\parallel}=d_{\parallel}E^{\prime}_{zz}=(2\pi)3.55(29)\epsilon$ GHz, 
$\Omega_{\perp}=d_{\perp}\sqrt{E^{\prime2}_{xx}+E^{\prime2}_{yy}}=(2\pi)13.8(8)\epsilon$ GHz\cite{ovartchaiyapong2014dynamic}. The observed (2$\pi$)2.19(14)~MHz/V$_{p}$ mechanically driven spin Rabi oscillation therefore corresponds to 1.59(14)$\times10^{-4}$ V$_{p}^{-1}$ in strain. 

The impedance of the micro-sized transducer is around 500~$\Omega$ at 3~GHz. To suppress power reflections from the impedance mismatch, we engineer the length of the RF transmission line on the circuit board to act as a $\lambda/4$ transformer. The actual voltage applied at the transducer is close to twice the voltage of that in a 50~$\Omega$ matched network. We note that the electrical power and voltage reported in the main text are quoted for a 50~$\Omega$ impedance matched load. The actual power applied to the transducer is less. Therefore, the power-to-strain efficiency of the device can be improved further using an impedance matching network.

\section{Potential quantum applications}

Even though the single spin-phonon cooperativity of the current device is low even at low temperature ($T=100$~mK), $\eta=2\pi \frac{g^{2}T_{2}}{\kappa \bar{n}}\sim10^{-7}$, where $g\sim 1$~Hz is the single spin-phonon coupling rate from zero-point motion of resonator (note that this is $10^{3}$ higher than HBAR devices made in the past\cite{macquarrie2013mechanical}), $T_{2}$ is the coherence time of an NV center spin, $\kappa=\omega_{m}/Q$ is the resonator damping rate, $\bar{n}$ is the phonon occupation number at temperature $T$. The ensemble spin-phonon coupling is enhanced relative to the single spin-phonon coupling by $\sim \sqrt{N}$, where $N$ is the number of coupled NV centers. For a dense ensemble of NV centers with density $10^{19}$~cm$^{-3}$\cite{choi2017depolarization}, there are close to $N=10^{10}$ addressable NV centers in a $(10 \mu\text{m})^{3}$ volume resonator. The effective spin-phonon coupling is $\sqrt{N}g\sim100$~kHz. The coupling can be further increased by considering alternative defect centers with stronger intrinsic spin-phonon coupling, i.e., SiV centers in diamond have been shown to have single spin-phonon coupling $\sim$100~THz/strain\cite{meesala2018strain}, around 5000 times larger than that for an NV center. A diamond SCHBAR with SiV centers at a moderate density of $10^{14}$~cm$^{-3}$ has ensemble spin-phonon coupling up to 1~MHz. Such high coupling can introduce a splitting in the resonator spectrum, which is detectable for a high quality resonator ($Q\sim10^{4}$, $\kappa\sim300$ kHz). When the resonator and the SiV center spin transition frequencies are detuned, the frequency pull of the resonator due to coupling to SiV ensemble can be used for dispersive read-out\cite{blais2004cavity} of spin state and for spin manipulation of the resonator state.


\providecommand{\latin}[1]{#1}
\makeatletter
\providecommand{\doi}
  {\begingroup\let\do\@makeother\dospecials
  \catcode`\{=1 \catcode`\}=2 \doi@aux}
\providecommand{\doi@aux}[1]{\endgroup\texttt{#1}}
\makeatother
\providecommand*\mcitethebibliography{\thebibliography}
\csname @ifundefined\endcsname{endmcitethebibliography}
  {\let\endmcitethebibliography\endthebibliography}{}

\end{document}